\documentclass[reprint, amsmath,amssymb, nobibnotes, aps, prb, superscriptaddress, showkeys, citeautoscript]{revtex4-2}

\usepackage[utf8]{inputenc}
\usepackage[T1]{fontenc}
\usepackage{orcidlink}
\usepackage[normalem]{ulem}
\usepackage[version=4]{mhchem}
\usepackage{bm}
\usepackage{braket}
\usepackage{siunitx}
\usepackage{bibunits}
\usepackage{amsmath}
\usepackage{jabbrv}
\usepackage{graphicx}
\usepackage[mathlines]{lineno}
\usepackage{booktabs}
\usepackage{xcolor}
\makeatletter
\@ifundefined{selectlanguage}{}{\renewcommand{\selectlanguage}[1]{}}
\makeatother
\DeclareMathAlphabet\mathbfcal{OMS}{cmsy}{b}{n}

\newcommand{\editorr}[2]{%
  \expandafter\newcommand\csname #1note\endcsname[1]{%
    \textcolor{#2}{(\textbf{#1:} ##1)}}%
  \expandafter\newcommand\csname #1\endcsname[1]{%
    \textcolor{#2}{##1}}%
  \expandafter\newcommand\csname #1cancel\endcsname[1]{%
    \textcolor{#2}{\sout{##1}}}%
  \expandafter\newcommand\csname #1change\endcsname[2]{%
    \textcolor{#2}{\sout{##1} ##2}}%
  \newenvironment{#1text}{\color{#2}}{\color{black}}
}

\editorr{LB}{red}

\newcommand\QE{\textsc{Quantum ESPRESSO}}

\newcommand\THEOSMARVEL{Theory and Simulation of Materials (THEOS), and National Centre for Computational Design and Discovery of Novel Materials (MARVEL), {\'E}cole Polytechnique F{\'e}d{\'e}rale de Lausanne, 1015 Lausanne, Switzerland}
\newcommand\PSI{Laboratory for Materials Simulations, Paul Scherrer Institut (PSI), 5232 Villigen, Switzerland}
\newcommand\BREMEN{U Bremen Excellence Chair, Bremen Center for Computational Materials 
Science, and MAPEX Center for Materials and Processes, Universit\"at Bremen, 28359 Bremen, Germany}
\newcommand\HMI{Hybrid Materials Interfaces Group, Faculty of Production Engineering, Universit\"at Bremen, Bremen, Germany}

\begin{document}

\begin{bibunit}[apsrev4-2]
\newcommand\TITLE{A grand potential for molten-salt electrochemical phase diagrams from first principles}
\title{\TITLE}

\author{{Lorenzo Bastonero}\orcidlink{0000-0001-9374-1876}} 
\email{lbastone@uni-bremen.de} \affiliation{\BREMEN}
\author{{Lucio Colombi Ciacchi}\orcidlink{0000-0003-1444-9733}} \affiliation{\BREMEN} \affiliation{\HMI}
\author{{Nicola Marzari}\orcidlink{0000-0002-9764-0199}} \affiliation{\BREMEN} \affiliation{\THEOSMARVEL} \affiliation{\PSI}

\begin{abstract}
The clean-energy transition is driving unprecedented demand for metals, whose extraction is among the largest industrial sources of carbon emissions.
Molten-salt electrolysis offers a path toward carbon-free production of aluminum, iron, titanium, and critical metals. However, corrosive operating conditions and complex, multicomponent feedstocks make experimental development costly and challenging.
Here, we introduce a first-principles framework for constructing electrochemical phase diagrams of multicomponent systems to accelerate electrode and electrolyte design and guide the selection of compatible materials and operating conditions for electrochemical cells.
We derive a grand potential that generates these diagrams at scale, and supply free energies using first-principles data obtained through a fine-tuned foundation machine-learned interatomic potential.
For silicon in molten \ce{CaCl2}, the construction not only recovers the experimental diagram when using the first-principles framework on a limited set of structures, but also predicts further domains when extended to a larger set of crystal structures.
Finally, we demonstrate the predictive power of the approach by screening ten alkali and alkaline-earth halide electrolytes for aluminum electrowinning. 
Three thermodynamic criteria identify \ce{NaF}, where cryolite emerges as the stable aluminum-bearing phase, recovering the chemistry that Hall and H{\'e}roult established empirically in 1886.
The framework provides a first-principles basis for selecting compatible electrodes, electrolytes, and operating conditions toward low-carbon metal production from complex source materials.
\end{abstract}

\flushbottom
\maketitle
\thispagestyle{empty}

\section*{Introduction}
The growing demand for metals to support clean-energy technologies intensifies the need for metal extraction~\cite{IEA2025}, which is itself a global driver of environmental change~\cite{Giljum2025}.
As an example, steelmaking alone emits about 7\% of the total global \ce{CO2} per year~\cite{Wang2023}.
In the last decades, molten-salt electrolysis has attracted renewed interest as a platform for cleaner metal production~\cite{Sadoway1995,Daehn2021,Daehn2024a}.
Inert, oxygen-evolving anodes enable metal extraction without direct carbon emissions and are reaching industrial demonstration in steel and aluminum production~\cite{Allanore2013,Du2023,Padamata2023}.
Furthermore, molten salts increasingly support the recycling of critical metals~\cite{Xi2020,Li2023} and the production of rare earths~\cite{Karati2025}, and are even employed in the electrolysis of lunar and Martian regolith for in situ oxygen and metal production~\cite{Meurisse2022,Zheng2022,Maes2024}.
To achieve green production in all these applications, with metal and oxygen as main by-products, the molten-salt electrochemical cells share several design constraints; for instance, the electrolyte must dissolve the feedstock, the anode must evolve oxygen without corroding, and the reduced metal must remain stable at the cathode.
Which phases are stable, and which dissolve, deposit, or corrode, at a given electrode potential, oxide-ion activity, temperature, and melt composition are therefore key questions of any molten-salt process and material design.

Pourbaix answered the analogous question for aqueous systems in the 1950s~\cite{Pourbaix1951}, and Littlewood devised the corresponding representation for molten chlorides a decade later~\cite{Littlewood1962}, mapping the stability of metals, salts, and oxides in the plane of electrode potential and oxide-ion activity.
Littlewood diagrams therefore remain the conceptual backbone of molten-salt corrosion and electroreduction, yet they describe the stability of only one metal element at a time in a given electrolyte, excluding the coupled equilibria of multiple metals under investigation. This restriction limits their use in systematically screening candidate electrolytes and electrodes. Moreover, their construction also remains largely manual, relying on the Nernst equation.
Recent work has automated Littlewood-diagram construction for molten chlorides through Gibbs-energy minimization with CALPHAD software~\cite{Brosh2024}. This approach can accommodate multiple metals, but its applicability depends on thermodynamic database coverage of all relevant phases. Another recent work has extended chloride stability maps to impurity variables while incorporating first-principles vibrational free energies~\cite{Ham2026}.
What is missing is an explicit thermodynamic potential for \emph{any} molten-salt \ce{MXz} electrochemical diagrams: an object that ranks arbitrary competing phases on a common free-energy scale and can be populated systematically from computational materials data.
In aqueous electrochemistry this step has been taken.
Free-energy formulations of Pourbaix diagrams turn a set of equilibrium lines into a thermodynamic landscape -- the Pourbaix potential -- that naturally accommodates additional contributions such as surface energies and connects to computational materials databases~\cite{Sun2019,Patel2019,Huang2019}.
The construction has since been placed within the general geometry of high-dimensional phase diagrams, where open systems are described by \emph{grand potentials} and predominance regions are obtained by projecting the lower envelope onto chemical-potential space~\cite{Sun2021,Chen2024,Chen2024a}. This formulation is geometrically dual to the more familiar convex-hull in composition space.
These grand-potential differences measure the thermodynamic competition between phases under specific external conditions~\cite{Wang2024}.

No such object has been derived for molten salts.
Moreover, when combining this formulation with first-principles data, the vibrational and thermal contributions that are commonly neglected at ambient temperature in Pourbaix diagrams~\cite{Persson2012} become important ingredients of phase stability at 700~K and above, i.e., the relevant temperature range for molten-salt applications.

Here we derive a generalized grand potential for binary molten-salt electrochemical equilibria whose natural variables are the electrode potential and the oxide-ion activity, and whose lower envelope reduces to Littlewood's original diagram for fused chlorides.
Because the potential is linear in both variables, we can easily determine the stability domains for any number of competing phases using a simple geometrical construction.
We first illustrate the construction on iron in molten \ce{LiCl}, using tabulated thermochemical data only: iron has recently been demonstrated as an oxygen-evolving anode in molten chlorides and carbonates~\cite{Du2023}.
We then present a comprehensive framework for computing these free energies from first principles for solids and gases, with liquids and dissolved species treated by controlled approximations; experimental thermochemistry enters only through element-wise shifts of the gas references, so that no system-specific experimental input is further needed.
This is achieved by combining density-functional theory (DFT) with vibrational free energies from density-functional perturbation theory (DFPT)~\cite{Baroni2001} and from a first-principle accurate fine-tuned foundation equivariant neural network interatomic potential, which allows us to extend the phase space to an arbitrary number of competing phases.
By this approach, we show that we recover and expand the experimental diagram of silicon in molten \ce{CaCl2} previously reported in Ref.~\onlinecite{Yasuda2007}.
Finally, we use the framework at scale on a practical electrolyte-engineering problem, screening ten alkali and alkaline-earth halides as candidate solvents for aluminum electrowinning.
Three thermodynamic criteria, read off the same set of diagrams, select \ce{NaF}, whose stable aluminum-bearing phase is cryolite itself. With this result, we recover from first principles the chemistry that Hall and H{\'e}roult arrived at empirically in 1886.

\section*{A grand potential for generalized Littlewood diagrams}
\begin{figure*}
    \centering
    \includegraphics[width=0.9\textwidth]{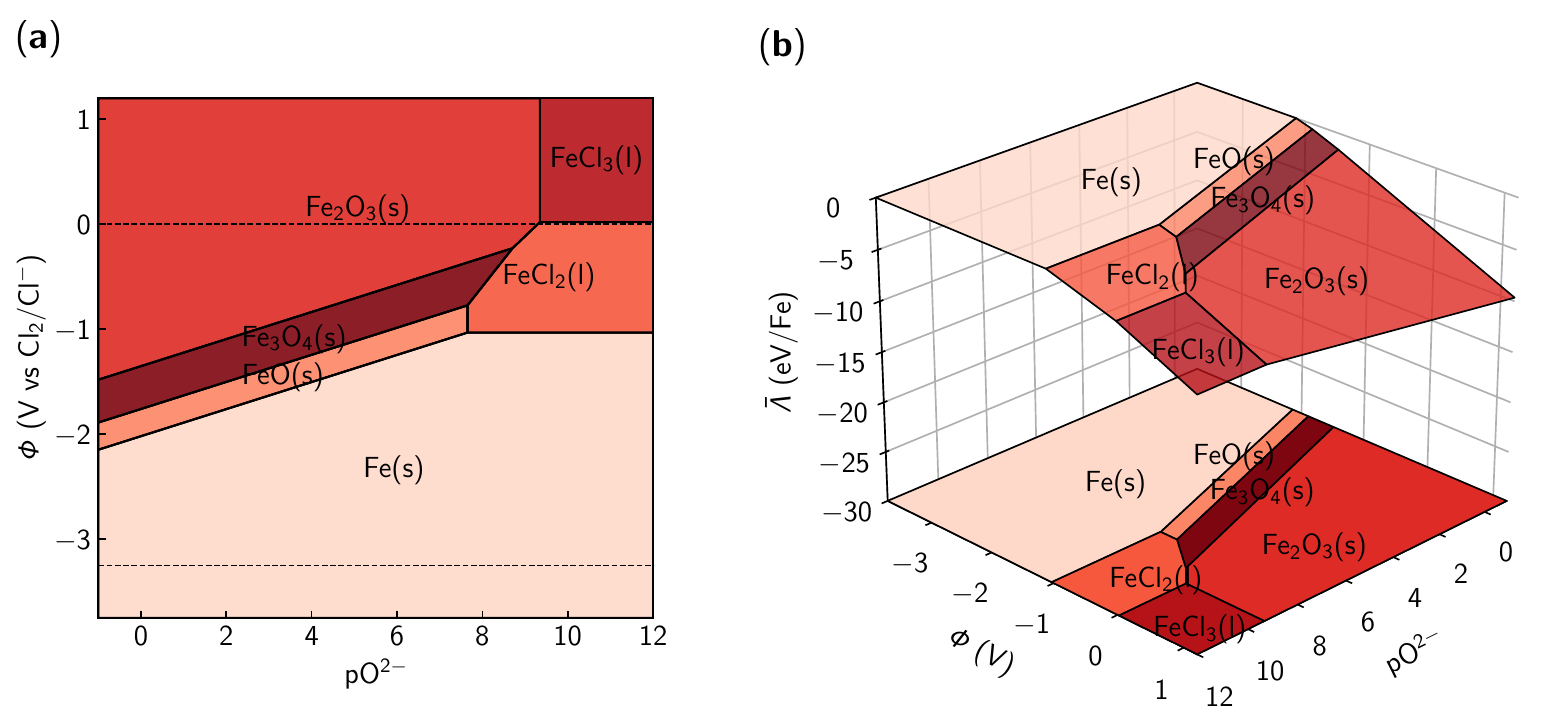}
    \caption{
        \textbf{Littlewood diagram of iron in molten \ce{LiCl} and its corresponding grand-potential concave landscape.}
        (a)~Stable Fe-bearing phases at 1300~K as a function of the oxide-ion activity, $\mathrm{pO^{2-}} = -\log a_{\ce{O^2-}}$, and the electrode potential $\phi$, referenced to the \ce{Cl2}/\ce{Cl-} couple.
        Formation Gibbs free energies are experimental values from the NIST-JANAF tables~\cite{Chase1998}.
        Dashed lines delimit the electrochemical stability window of the melt, between chlorine evolution ($\phi = 0$) and lithium deposition ($\phi = G^f_{\ce{LiCl}}$).
        (b)~The same equilibrium as a free-energy landscape: every competing phase contributes a plane $\overline{\Lambda}_i(\phi, \ce{pO^{2-}})$, and the facets of their lower envelope (upper surface) are the stability domains, whose projection onto the $(\phi, \ce{pO^{2-}})$ plane (lower surface) is panel~(a).
        Both panels follow from a geometric construction, with no equilibrium reaction enumerated by hand.
    }
    \label{fig:fe}
\end{figure*}
Littlewood diagrams~\cite{Littlewood1962}, introduced for fused chlorides (\ce{MCl2}), are drawn in the plane of the electrode potential $\phi$, referenced to the \ce{Cl2}/\ce{Cl-} couple, and of the oxide-ion activity $a_{\ce{O^2-}}$, reported as $\ce{pO^{2-}} = -\log_{10} a_{\ce{O^2-}}$.
The two variables encode the electrochemical driving force and the basicity of the melt, respectively, and together determine the stability of dissolved ions, salts, metals, and precipitated oxides.
However, the classical construction based on the Nernst equation has no free-energy axis.
Such an axis would turn the diagram into a thermodynamic landscape, where all competing phases are directly comparable to one another, and further contributions can be added, such as surface energies~\cite{Sun2019}.

We add this axis, and at the same time generalize the construction from chlorides to an arbitrary binary molten salt \ce{MX_z}, of melt cation \ce{M^{z+}} and halide anion \ce{X-}, by devising a grand potential for solvated ions and solid precipitates in open exchange with a reservoir of molten salt and oxide ions.
Its control variables are the temperature, the electrode potential, \ce{pO^{2-}}, and the activities of the dissolved species, which we approximate by their concentrations relative to the standard state.
To obtain what we will refer to as the generalized Littlewood grand potential, we perform a Legendre transformation of the Gibbs free energy with respect to the numbers of melt cations, halogen and oxygen atoms, and elementary charges exchanged with the reservoir, subject to the molten-salt--halogen and cation-oxide equilibria.
We report the derivation in the Methods.
Normalized by the number $N^*$ of solute atoms in the phase, the potential $\overline{\Lambda}$ reads
\begin{equation}
    \begin{aligned}
        \overline{\Lambda} =
        \frac{1}{N^*}
        &\left \{ ~[G^f - N_{\ce{M}}\mu_{\ce{MX_z}} - \frac{N_{\ce{O}}}{z}(\mu_{\ce{M_2O_z}} - 2\mu_{\ce{MX_z}})] \right .
        \\
        &+[z N_{\ce{M}} - N_{\ce{X}}  - 2 N_{\ce{O}} - q] \phi
        \\
        & \left . +\ln(10) k_B T N_{\ce{O}} \ce{pO^{2-}} \right \}
        ~,
    \end{aligned}
    \label{eq:littlewoodpotentialexplicit}
\end{equation}
where $G^f$ is the formation Gibbs free energy of the phase, $\mu_A$ the chemical potential of specie $A$, $N_i$ its stoichiometric coefficient of species $i$, $q$ its net charge in units of the elementary charge, and $k_B$ the Boltzmann constant; all energies and chemical potentials are per atom in eV, so that the numerical value of $\phi$ in volts gives the corresponding electron energy in electronvolts.
The zero of the grand potential is set by the reservoir and by the elemental standard state of the solute: $\overline{\Lambda}_i = 0$ marks equilibrium of phase $i$ with the pure solute at unit activity under the imposed melt conditions, so that $\overline{\Lambda}_i < 0$ signals its spontaneous formation from the element, while equilibrium between two competing phases, including the solvated species, is the boundary $\overline{\Lambda}_i = \overline{\Lambda}_j$. For a single solute element, the stable phase at given $(\phi, \ce{pO^{2-}})$ is the one with the lowest $\overline{\Lambda}_i$.

Because $\overline{\Lambda}$ is linear in both $\phi$ and \ce{pO^{2-}}, every competing phase is a plane in a three-dimensional free-energy space, and the phase diagram is the projection of the lower envelope of these planes, i.e., the concave surface.
Introducing a lifted coordinate $\lambda$, each phase defines the \emph{half-space}
\begin{equation}
    \lambda - \overline{\Lambda}_i(\phi,\ce{pO^{2-}})
    = \lambda-a_i\phi-b_i\ce{pO^{2-}}-c_i \le 0
    ~,
    \label{eq:halfspace}
\end{equation}
with $a_i$, $b_i$, and $c_i$ the coefficients of its plane, and the intersection of all half-spaces is the lower concave envelope whose facets project onto the predominance regions~\cite{Chen2024a}.
Thanks to this geometric construction, pairwise equilibrium lines are never enumerated to draw the stability map of a single metal.

The construction extends to several solute elements, as in an alloy, where the lowest single-phase plane is no longer sufficient. In fact, at fixed solute composition the stable state is in general a mixture of phases, requiring one mass balance per element.
Since $\overline{\Lambda}$ is linear in the stoichiometric coefficients, any composition-weighted combination of phases is itself a plane, and it enters Eq.~\ref{eq:halfspace} as one more half-space.
As a consequence, we can minimize over these combinations, and by relying on the implementation and algorithms introduced in Ref.~\cite{Patel2019} for many-element Pourbaix diagrams, our implementation therefore carries the construction over to multicomponent solutes (see also Methods).

Figure~\ref{fig:fe} illustrates the construction for iron in molten \ce{LiCl} at 1300~K, using experimental thermochemical data~\cite{Chase1998}.
Panel~(a) is the familiar Littlewood predominance map: metallic iron spans the reducing part of the electrochemical window, the oxides \ce{FeO}, \ce{Fe3O4}, and \ce{Fe2O3} follow in sequence with increasing potential on the basic side, and the molten chlorides \ce{FeCl2} and \ce{FeCl3} take over the acidic corner.
A single diagram thus separates immunity, passivation, and active dissolution, the three regimes that govern the corrosion of structural alloys in chloride melts.
Panel~(b) shows the geometric origin of these regions: each phase is a plane in the space spanned by $\phi$ and \ce{pO^{2-}}, and the diagram is the projection of the facets of their lower intersection envelope.
We note that in the passivation regime we find pure iron, which was discovered as an oxygen-evolving anode in Ref.~\cite{Du2023}.
The lithium ferrite scale that protects such an anode~\cite{Du2023} is however absent from the diagram, since no \ce{Li}--\ce{Fe}--\ce{O} phase is included in the NIST-JANAF database used here, highlighting the need of larger thermochemistry data that the following sections address.

\section*{First-principles framework}
Turning generalized Littlewood diagrams into a predictive tool requires consistent Gibbs free energies for all competing phases, which we compute from first principles following a similar strategy introduced for computational Pourbaix diagrams~\cite{Persson2012}.
A key difference is that, at the high temperatures relevant for molten-salt applications, neither the thermal contributions nor the volatile species, such as monatomic and diatomic gases, can be neglected.
In the following, we summarize the main ingredients of the approach, and refer to the Methods for details.

For crystalline phases we combine DFT ground-state energies with harmonic vibrational free energies, which we compute either from density-functional perturbation theory (DFPT) or by finite differences with NequIP-OAM-MAD-L, a foundation machine-learned interatomic potential (MLIP) that we fine-tuned to the same exchange-correlation functional approximation (i.e., PBEsol) used for the DFT and DFPT references.
Gases are treated using the ideal gas limit.
For simplicity, liquid and solvated-ion free energies are not computed explicitly in this first implementation: every condensed phase, including the molten salt and any solute phase that is liquid at the temperature of the diagram, such as metallic aluminum above 933~K, is approximated by the harmonic free energy of its lowest-energy crystal, while the standard chemical potential of a dissolved ion is referenced to the stoichiometrically corresponding halide (as already practiced in the original work of Littlewood~\cite{Littlewood1962}). For dissolved species, $\mu_A=\mu_A^\circ+k_BT\ln a_A$, where $\mu_A^\circ$ is the standard chemical potential and $a_A$ the activity of species $A$, approximated in the dilute limit by its concentration relative to the standard state.

The crystal approximation for liquids neglects the entropy of fusion, approximately $\Delta S_\mathrm{fus}(T-T_m)$, i.e. about 0.04~eV/atom for aluminum at 1300~K, using the enthalpy of fusion of the NIST-JANAF tables~\cite{Chase1998}.
Since these free energies set the electrochemical stability window of the diagrams, we show in Supplementary Figure 1 that this approximation reproduces, to a good extent, the experimental trends through the melting points.

The solid competing phases are drawn from a curated experimental database of inorganic crystals~\cite{Huber2026} rather than from the compounds selected for a hand-drawn diagram, as broader coverage can add stability domains or reorder the accessible phases.
In the applications presented in the next sections, all structures are retained in the construction of the diagrams, including those whose harmonic phonon spectrum contains imaginary modes (see Methods).
Their construction restricted to dynamically stable structures are reported in the Supplementary Information, and commented in the Supplementary Note~4.
Since DFPT for every candidate rapidly becomes prohibitive, larger structure sets rely on NequIP-OAM-MAD-L, which we obtain by fine-tuning the NequIP-OAM-L foundation model~\cite{Kavanagh2026}, built on the NequIP architecture~\cite{Batzner2022,Tan2025}, on the PBEsol MAD dataset~\cite{Mazitov2025a}.
It supplies only the vibrational term, and reproduces the DFPT vibrational free energies with a mean absolute error of 12.7~meV/atom at 1000~K (see Supplementary Notes and Supplementary Figure 3).

Similarly to Ref.~\onlinecite{Persson2012}, we apply rigid shifts of the elemental halogen and oxygen free energy references to correct for the systematic errors of semi-local DFT for molecular energies~\cite{Perdew1981,Hammer1999,Wang2006}.
Molten-salt equilibria, however, span a much wider temperature range than the ambient conditions of Pourbaix diagrams, so we fit one temperature-independent shift per element to experimental formation free energies over 300--1500~K (see Methods and Supplementary Note 1).
The shifts are applied to condensed phases only, since corrections fitted on ionic salts are not transferable to covalent molecules~\cite{Wang2006} (see also Supplementary Note~1).

We note that, whenever available, a shift of the oxide-ion standard chemical potential may be used to align the calculated cation-oxide solubility limit with experiment (as done in the next section to compare our results with those of Ref.~\onlinecite{Yasuda2007}).
Such a shift is a rigid translation of the \ce{pO^{2-}} axis, which moves the origin of the scale without altering the diagram; for that comparison it amounts to $+0.21$~eV, i.e. 0.94 units of \ce{pO^{2-}} at 1120~K, the standard-state difference between crystalline \ce{CaO} and the \ce{CaO} solubility of 0.111 anion fraction adopted in the experimental construction.

\begin{figure*}[!ht]
    \centering
    \includegraphics[width=0.9\textwidth]{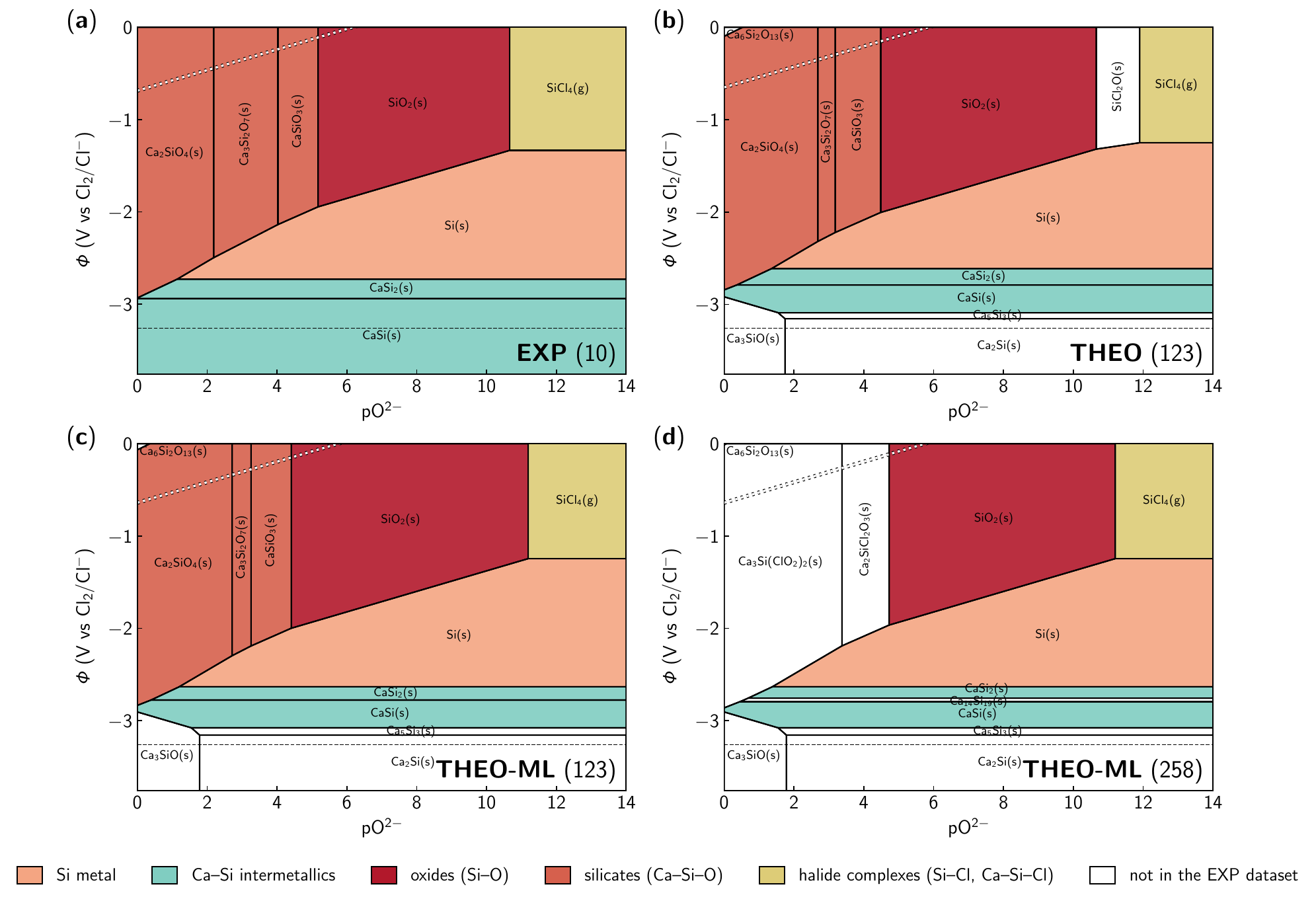}
    \caption{
        \textbf{Experimental versus theoretical Littlewood diagrams of silicon in molten \ce{CaCl2}.}
        Stable Si-bearing phases as a function of the oxide-ion activity, $\mathrm{pO^{2-}} = -\log a_{\ce{O^2-}}$, and the electrode potential $\phi$, referenced to the \ce{Cl2}/\ce{Cl-} couple.
        The number in parentheses counts the phases entering the construction: for (a), the species of the experimental construction; for (b)--(d), the distinct Si-bearing crystal structures considered (including polymorphs), to which gaseous \ce{SiCl4} is added in all three panels.
        (a)~Experimental data from Ref.~\onlinecite{Yasuda2007} ($T = 1123$~K).
        (b)--(d)~First-principles free energies within the framework described in this work ($T = 1120$~K), combining DFT (PBEsol) electronic energies with harmonic vibrational contributions from (b)~DFPT and (c)--(d) NequIP-OAM-MAD-L.
        In particular, for (c)~the subset of structures shared with the DFPT dataset and (d)~the full dataset.
        Domains are color-coded by species category (bottom legend); white marks phases absent from the experimental dataset of~(a).
        Black dashed lines delimit the electrochemical stability window of the melt, between chlorine evolution ($\phi = 0$) and calcium deposition ($\phi = G^f_{\ce{CaCl2}}/2$).
        The white dotted line marks the oxidation of dissolved oxide (\ce{2 O^2- -> O2 + 4 e-}).
    }
    \label{fig:yasuda}
\end{figure*}

\section*{Recovering and expanding experimental diagrams}
We assess the accuracy, robustness, and capability of the framework against the experimental reference diagram for the \ce{Ca}--\ce{Cl}--\ce{O}--\ce{Si} system reported in Ref.~\onlinecite{Yasuda2007}.
The comparison is shown in Fig.~\ref{fig:yasuda}: Fig.~\ref{fig:yasuda}(a) shows the experimental data, while Fig.~\ref{fig:yasuda}(b)--(d) display the calculated diagrams with increasing phase-space coverage. 
Coverage is counted in distinct crystal structures rather than stoichiometries, so that polymorphs compete individually.
Fig.~\ref{fig:yasuda}(b) uses DFPT vibrational free energies for the 123 Si-bearing structures of our DFPT dataset, against the ten species of the experimental construction. Fig.~\ref{fig:yasuda}(c) repeats the construction with NequIP-OAM-MAD-L vibrational free energies on the same 123 structures. Finally, Fig.~\ref{fig:yasuda}(d) extends it to all 258 Si-bearing structures of the inorganic crystal dataset.

For all three constructions, the major stability domains and their adjacency sequence across the electrochemical window are recovered, even considering the simplicity of the harmonic approximation to calculate the free energies.
The DFPT construction of Fig.~\ref{fig:yasuda}(b) demonstrates two points.
First, generalized Littlewood diagrams assembled from \emph{ab initio} solid-state thermodynamics recover the experimental sequence of silicates, silica, silicon, and \ce{Ca}--\ce{Si} alloys across the electrochemical window.
Second, even a modest expansion of the phase space already refines the diagram: the white domains mark phases that were not considered in the experimental construction, namely the \ce{Ca5Si3} and \ce{Ca2Si} alloys near the cathodic limit, the inverse perovskite \ce{Ca3SiO} under strongly reducing, basic conditions, a narrow \ce{Ca6Si2O13} domain at the anodic limit, and \ce{SiCl2O} between \ce{SiO2} and \ce{SiCl4}(g).
The latter two are however dynamically unstable in DFPT and disappear once only stable structures are considered (Supplementary Figure~4).

The accessible phase space nevertheless remains limited by the cost of DFPT.
Before expanding the phase space, we verify that the fine-tuned potential can replace DFPT without altering the diagram.
Fig.~\ref{fig:yasuda}(c) repeats the construction at fixed structure set, with the DFPT vibrational free energies replaced by those of NequIP-OAM-MAD-L.
Notably, the diagram remains essentially unchanged, with marginal boundary shifts, and with only the \ce{SiCl2O} domain not reproduced, which is in any case dynamically unstable also using the MLIP.
In the Supplementary Information we further compare and discuss the dynamical stability predictions between DFTP and MLIP.
This qualitatively validates the ML potential directly at the level of the phase diagram, complementing the benchmarks reported in the Supplementary Information.

We then expand the phase space to all 258 Si-bearing structures in the dataset, Fig.~\ref{fig:yasuda}(d), including 135 for which systematic DFPT would be computationally demanding.
A \ce{Ca}--\ce{Si} intermetallic band appears near the cathodic limit, between \ce{CaSi2} and \ce{CaSi}, and is assigned to \ce{Ca14Si19}.
Experimentally, this phase decomposes into \ce{Ca3Si4} and \ce{CaSi2} on cooling below 1163--1180~K~\cite{Moll2017}, so at 1120~K the two-phase mixture is expected.
Both phases are included in the 258-structure set, and our free energies reproduce the ordering, with \ce{Ca14Si19} the high-temperature phase, but place the decomposition near 700~K. 
Electronic DFT energies alone favor the mixture at all temperatures, and it is the vibrational term that stabilizes \ce{Ca14Si19}, consistent with the role of lattice dynamics noted in Ref.~\onlinecite{Moll2017}.
At 1120~K, the free energy of the decomposition \ce{2 Ca14Si19 -> 9 Ca3Si4 + CaSi2} is only $+0.6$~meV/atom in favor of \ce{Ca14Si19}, an order of magnitude below the spread between MLIP and DFPT vibrational free energies.
Resolving such a small free-energy difference at this temperature, independently of the interatomic potential, requires controlling several further contributions: thermal expansion, anharmonicity, and the exchange-correlation functional approximation itself.
Importantly, in this case the question affects only the label of that domain: if \ce{Ca14Si19} is excluded from the competing phases, \ce{Ca3Si4} occupies exactly the same region of the diagram, and no other boundary moves by more than 0.01 units of \ce{pO^{2-}}.
Neither phase appears in the experimental construction nor in the DFPT panel of Fig.~\ref{fig:yasuda}(b), so both the domain and its degeneracy are revealed only by the broader phase space.
At low \ce{pO^{2-}}, two quaternary oxychlorides, \ce{Ca3Si(ClO2)2} and \ce{Ca2SiCl2O3}, absent from both the experimental and the DFPT constructions, become stable over substantial regions, replacing part of the silicate sequence.
Both have since the work of Yasuda and coworkers~\cite{Yasuda2007} been identified by X-ray diffraction, as the intermediate phases that form when \ce{Ca2SiO4} and \ce{CaSiO3} dissolve in molten \ce{CaCl2} between 1047 and 1233~K~\cite{Cheng2022}.
Of the two, \ce{Ca2SiCl2O3} is dynamically stable according to the MLIP phonons, whereas \ce{Ca3Si(ClO2)2} shows a soft mode.
The softening may be a physical instability, reflecting thermal fluctuations at high temperature that the harmonic approximation cannot capture, or an error of the potential.
Since the phase is observed experimentally, we regard its domain as robust, and leave the verification, which requires DFPT or anharmonic lattice dynamics on a 40-atom unit cell, to future work.

\section*{Rationalizing the Hall--H{\'e}roult electrolyte}
\begin{figure*}[!t]
    \centering
    \includegraphics[width=1.0\textwidth]{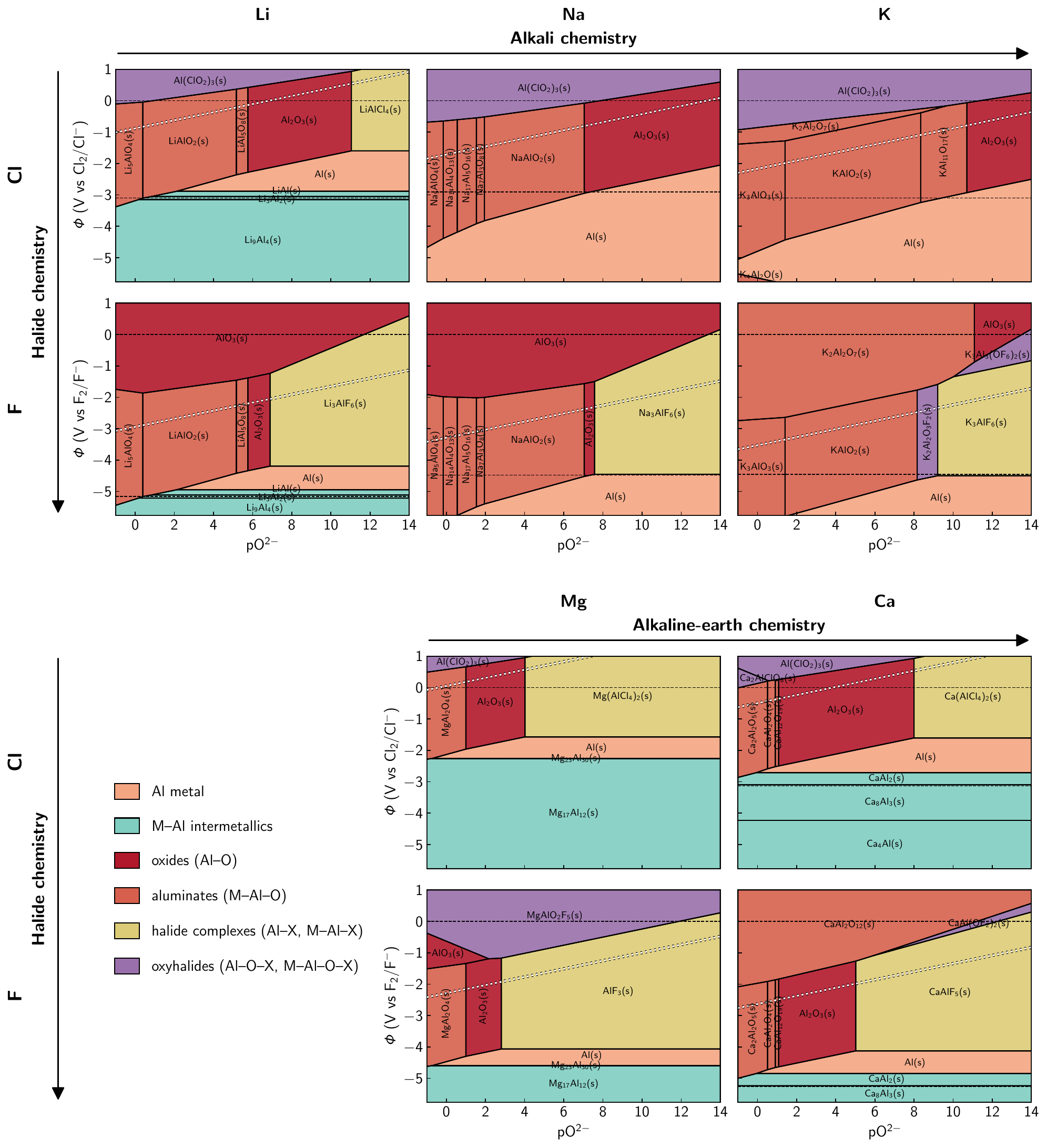}
    \caption{
        \textbf{Computed generalized Littlewood diagrams comparing aluminum in molten alkali and alkaline-earth halides at 1300~K.} 
        Stable Al-bearing phases as a function of the \ce{pO^{2-}} and $\phi$, referenced to the \ce{X2}/\ce{X-} couple of each melt. 
        Each panel corresponds to the \ce{MX_$z$} melt (top block: alkali melts, M = Li, Na, K; bottom block: alkaline-earth melts, M = Mg, Ca, with columns aligned by period; within each block, rows correspond to X = Cl, F).
        Free energies combine DFT electronic energies with harmonic vibrational contributions from the fine-tuned MLIP (see main text).
        Black dashed lines delimit the electrochemical stability window of the melt.
        The white dotted line marks the oxidation of dissolved oxide (\ce{2 O^{2-} -> O2 + 4 e-}). 
        Domains are color-coded by species category (bottom-left legend).
    }
    \label{fig:littlewood_al_ml_grid}
\end{figure*}
Having assessed the accuracy of the framework, we can now use it at large scale to rationalize the Hall--H{\'e}roult electrolyte for smelting aluminum, a prototypical electrolyte-engineering problem of molten-salt electrometallurgy.
In this process, alumina is dissolved in molten cryolite, \ce{Na3AlF6}, and electrolyzed in a reduction cell operating near 1230~K, about half the melting temperature of alumina.
Cryolite was identified as the solvent in 1886, and the bath composition, with \ce{AlF3} in excess of the cryolite stoichiometry and further additions such as \ce{CaF2}, has been refined over more than a century of practice.

If we were to choose a molten-salt composition for this process today, we would pose the choice as an electrolyte screening problem.
While several of its criteria concern transport properties, such as ionic and electrical conductivity, among others (see, e.g., Ref.~\onlinecite{Esmaily2020}), Littlewood diagrams address the fundamental thermodynamic constraints on the solvent: the melt should dissolve as much of the oxide source materials as possible, its electrochemical window should be wide enough to reduce the metal before the solvent itself decomposes, and it should yet leave metallic aluminum stable at the cathodic limit without alloying with the solvent metal cation.
The first constraint fixes the \ce{pO^{2-}} at which the dissolved aluminum-bearing species precipitate as an oxide; the lower this value, the more oxide the melt can dissolve.

Figure~\ref{fig:littlewood_al_ml_grid} compares generalized Littlewood diagrams for aluminum, computed with DFT formation energies and NequIP-OAM-MAD-L vibrational free energies, for ten alkali and alkaline-earth chlorides and fluorides at 1300~K.
While at this common temperature \ce{MgF2} and \ce{CaF2} are still solid, melting at 1536~K and 1691~K, the grid still lets us compare different chemistries on the same footing. 
In fact, since their free energies vary steadily with temperature, the phase ordering in the two panels would not change qualitatively at the higher temperatures where the salts are actually molten (see also Supplementary Figure 1).
We remark that, at the chosen temperature, these two panels lack a realizable process condition, rather than a well-defined thermodynamic reference.

Before examining the individual melts, two general trends emerge across the grid.
Across the halides, fluoride complexes aluminum more strongly than chloride: fluoroaluminate phases are stabilized over a large share of every fluoride electrochemical window, whereas chloroaluminates in the alkali chlorides are either not stabilized or confined to a small acidic corner of the \ce{LiCl} panel.
Moreover, fluorides show larger electrochemical stability, meaning a fluoride melt tolerates a larger potential range before the solvent itself decomposes.
Within each halide family, complexation then weakens down the group, from \ce{LiF} to \ce{KF} and from \ce{MgCl2} to \ce{CaCl2}.
These two trends govern the first two criteria mentioned above: low \ce{pO^{2-}} at which Al precipitates and wide electrochemical window of the melt, making fluoride melts already more appealing than chlorides, consistent with the low alumina solubility measured in chloride melts and with its increase upon fluoride addition~\cite{Picard1980}.
The third criterion follows no systematic trend: alloy formation depends on whether the melt cation forms an intermetallic with aluminum. When it does, the intermetallic truncates the metallic-aluminum stability domain at the cathodic limit.
A suitable molten salt should therefore sit high on the two trends and, at the same time, form no aluminum intermetallic.

The alkaline-earth melts are less successful on this last point.
The \ce{CaCl2} and \ce{MgCl2} melts stabilize extended \ce{Ca(AlCl4)2} and \ce{Mg(AlCl4)2} domains, so these are the chloride melts that do dissolve aluminum once sufficiently acidic, and they have been investigated in previous work~\cite{Yan2009}.
In both, however, the metallic-aluminum domain is truncated at the cathodic limit by \ce{Ca}--\ce{Al} and \ce{Mg}--\ce{Al} intermetallics, so the product may remain contaminated by the melt cation, as also reported experimentally in Ref.~\onlinecite{Yan2009}.
\ce{MgCl2} is the weaker of the two on the second criterion as well, its window being the narrowest of the chlorides.
The fluorides counterpart, \ce{CaF2} and \ce{MgF2}, are truncated in the same way, and are in any case solid at the operating temperature.

Among the alkali fluorides, the trends favor \ce{LiF}, which carries the most extended fluoroaluminate domain, but its aluminum domain is truncated by \ce{Li}--\ce{Al} intermetallics.
\ce{KF} forms no aluminum intermetallic and keeps the domain intact, yet sits lower on the complexation trend.
\ce{NaF} is the only composition that dissolves the feed over a wide range, spans a wide window, and leaves metallic aluminum untruncated.
However, the \ce{Na3AlF6}/\ce{Al} boundary is very close to the sodium-deposition line, so that the cathodic margin is small and the process is intrinsically prone to co-depositing sodium with aluminum.
Sodium co-deposition is therefore an intrinsic thermodynamic trait of the \ce{NaF} chemistry.
In fact, for industrial operation, additional \ce{AlF3} mitigates this limitation by shifting the activities~\cite{Ratvik2022}.

Read from the generalized Littlewood diagrams, the three thermodynamic criteria thus narrow ten candidate compositions to the one closest to what industry currently uses.
We note that the activities of the solvated species, including the oxide ion, are not computed here; their standard chemical potentials are referenced to the corresponding halides and oxides, and ideal dilution is assumed.
As stated earlier, a complete screening should also weigh properties such as ionic conductivity, density, viscosity, and cost.
In fact, \ce{KF}--\ce{AlF3} melts remain under investigation as low-temperature electrolytes for smelting aluminum~\cite{Padamata2023}, and other combinations, beyond the ten single-salt compositions considered here, remain to be explored.
The thermodynamic part can nevertheless be done first, for any metal, from the same three readings of a single diagram.

\section*{Discussion}
We have derived a grand potential for molten-salt electrochemical equilibria with electrode potential and oxide-ion activity as its natural variables. This gives the Littlewood construction an explicit free-energy axis and allows further thermodynamic contributions to be incorporated as additional terms in the potential.
The accompanying first-principles framework computes the Gibbs free energies of competing phases -- the main inputs to the grand potential -- using controlled approximations. These free energies are calculated using a fine-tuned foundation MLIP that accurately reproduces first-principles vibrational free energies of solids (MAE of 12.7~meV/atom at 1000~K), enabling diagram calculations to scale to a large number of competing phases.
We first demonstrated the usefulness of expanding the phase space in the O--Si--Ca--Cl system, where we recovered, from first principles, two recently detected phases and found both to be stable in the resulting diagrams.
Second, we showed an application to electrolyte engineering, qualitatively recovering the chemistry established empirically by Hall and H{\'e}roult more than a century ago for aluminum extraction.

A key strength of the present framework is its straightforward extension to electrochemical stability predictions for mixed-salt electrolytes. Each additional cation or anion in the reservoir can be readily introduced as an additional term in the Legendre transform, with its chemical potential determined by the activity of the corresponding salt component. The oxide-ion standard state can be defined with reference to the oxide of a chosen cation, naturally the one whose oxide precipitates first.
Importantly, the entire formulation is independent of the source of the free energies. CALPHAD assessments or experimental tabulations can therefore replace or complement first-principles values, as already practiced for Pourbaix diagrams~\cite{Persson2012}.

As foundation ML potentials become more and more accurate, two natural extensions could improve the first-principle thermodynamic description: a more complete treatment of solid-state free energies and explicit free energies for liquids and solvated species. 
For solids, anharmonicity and thermal expansion may be, for example, introduced using the stochastic self-consistent harmonic approximation~\cite{Errea2014,Monacelli2021a}, which treats these contributions non-perturbatively, and has been recently coupled with active learning of MLIPs demonstrating great efficiency and accuracy in predicting thermodynamics and phase transitions of crystals~\cite{Bastonero2026,Belli2025}.
For liquids and solvated species, free energies could be computed using thermodynamic integration and adiabatic substitution techniques~\cite{Frenkel2011}, which have already been combined with MLIPs and enabled calculations of molten-salt chemical potentials and melting points~\cite{Shi2022,Shah2023,Gibson2025,Shen2025}.
The growing availability of workflows for automated thermodynamic calculations~\cite{Menon2021,Menon2024,Bastonero2026}, accurate determination of ground-state energetics in transition-metal chemistries~\cite{Bastonero2025a,Ponet2024,Macke2024}, and active learning of MLIPs~\cite{Bidoggia2026,Liu2025} will enable accurate free-energy predictions for all relevant phases -- solids, liquids, solvated ions, and gases -- at scale, broadening the capabilities of the present framework.

Finally, we envision that coupling the proposed generalized Littlewood diagrams with first-principles predictions of diffusion coefficients, viscosity, and electrical conductivity will accelerate the discovery and design of electrode--electrolyte pairs for molten-salt electroreduction at operating temperatures and oxide-ion activities. Introducing surface free energies to the grand potential would further extend the framework to interfacial stability, providing a thermodynamic basis for investigating corrosion and electrodeposition.

This work represents therefore an important step towards the design and discovery of materials and processes for  electrolysis, and a key starting point toward green metal production from low-grade sources.

All data and code presented in this work are openly available, hoping this will support broad community adoption and accelerate advances in green electrometallurgy and related fields.

\section*{Methods}

\subsection*{Grand-potential analytical derivation}

Our derivation is conceptually inspired by the free-energy formulation introduced for aqueous Pourbaix diagrams by Sun and coauthors~\cite{Sun2019}.
The central idea is to endow the electrochemical phase diagram with an explicit free-energy axis, so that stability is determined by a thermodynamic lower envelope rather than only by pairwise equilibrium lines.
In the molten-salt case, this requires formulating a grand potential for solvated ions and solid precipitates in open exchange with a generic molten-salt reservoir \ce{MX_z}, at fixed \ce{pO^{2-}}, electrode potential, and dissolved-ion concentration.
Because these variables define the natural thermodynamic setting of generalized Littlewood diagrams, we refer to the resulting quantity as the \textit{generalized Littlewood grand potential}.

We construct the generalized Littlewood grand potential, $\Lambda$, by performing a Legendre transformation of the Gibbs free energy with respect to the exchanged numbers of cations, halogens, oxygen atoms, and elementary charges:
\begin{equation}
    \Lambda = 
    G - N_{\ce{M}}\mu_{\ce{M}}
      - N_{\ce{X}}\mu_{\ce{X}}
      - N_{\ce{O}}\mu_{\ce{O}}
      - q\phi
    ~,
    \label{eq:legendretransform}
\end{equation}
where $N_i$ is the number of particles of species $i$ and $q$ is the net charge in units of the elementary charge.
All chemical potentials and free energies are per particle and expressed in electronvolts; with this convention, $q\phi$ is in electronvolts when the numerical value of $\phi$ is given in volts.
The grand potential is then constrained to molten-salt--halide and cation-oxide equilibria, translating into the following equilibrium conditions
\begin{align}
    \ce{MX_z} &\rightleftharpoons \ce{M^{z+}} +z \ce{e^-} + \frac{z}{2}\ce{X_2}~, \label{eq:mxneq}\\
    \ce{M_2O_z} &\rightleftharpoons 2\ce{M^{z+}} + z\ce{O^{2-}}~. \label{eq:m2ozeq}
\end{align}
By writing the electrochemical potentials in solutions and using the first condition, we obtain the following relationships
\begin{align}
    \mu_{\ce{M}} &= \mu_{\ce{M^{z+}}} - z \phi = \mu_{\ce{MX_z}}^{\circ} - z \phi~, \label{eq:mum}\\
    \mu_{\ce{X}} &= \mu_{\ce{X^-}} + \phi = +\phi~, \label{eq:mux}
\end{align}
where $\mu_{X}^{\circ}=\mu_{X^-}^{\circ}=0$ represents the standard halogen electrode, and we assumed unit activity for both the halogen and the cation in the molten salt, as in Refs.~\onlinecite{Littlewood1962,Brosh2024}.
We can now write the electrochemical potential of oxygen ion in terms of the oxide-ion, using $\ce{pO^{2-}} = -\log a_{\ce{O^{2-}}}$, as
\begin{equation}
    \mu_{\ce{O}} = \mu_{\ce{O^{2-}}} +2\phi = \mu_{\ce{O^{2-}}}^{\circ} -\ln(10) k_B T \ce{pO^{2-}} + 2 \phi~.
    \label{eq:muo}
\end{equation}
To define the standard state of the oxide ion, we can use the second relationship, Eq.~\ref{eq:m2ozeq}, corresponding to the convention adopted by Littlewood~\cite{Littlewood1962}, to write
\begin{equation}
    \mu_{\ce{M_2O_z}} = 2~\mu_{\ce{M^{z+}}} + z \mu_{\ce{O^{2-}}}
    ~,
\end{equation}
and derive the relationship
\begin{equation}
    \mu_{\ce{O^{2-}}}^{\circ} = \frac{1}{z} (\mu_{\ce{M_2O_z}}^{\circ} - 2~\mu_{\ce{MX_z}}^{\circ})
    ~.
    \label{eq:muoxideion}
\end{equation}
where we substituted $\mu_{\ce{M^{z+}}}^{\circ}$ with $\mu_{\ce{MX_z}}^{\circ}$ from Eq.~\ref{eq:mum}.

The definition of this standard state allows us to draw the equilibrium
\begin{equation}
    2\ce{e^-} + \frac{1}{2} \ce{O_2} \rightleftharpoons \ce{O^{2-}}
    ~,
\end{equation}
using the relationship
\begin{equation}
    \phi = \phi_{\ce{O^{2-}}}^\circ
    + \frac{k_BT}{4} \ln \left(\frac{p_{O_2}}{p^\circ}\right)
    + \frac{\ln(10)k_BT}{2}\ce{pO^{2-}}
    ~,
\end{equation}
where $\phi^{\circ}_{\ce{O^{2-}}} = -\mu^{\circ}_{\ce{O^{2-}}}/2$ when the standard oxygen-gas formation chemical potential is set to zero, in the per-particle electronvolt convention adopted here.

The definition of the standard states in molten salts allow us to determine the final form of the grand potential.
By substituting the relationships Eqs.~\ref{eq:mum}, \ref{eq:mux}, \ref{eq:muo}, and \ref{eq:muoxideion} in Eq.~\ref{eq:legendretransform}, and by rearranging the terms, we obtain the generalized Littlewood potential which yields Eq.~\ref{eq:littlewoodpotentialexplicit}.
We remind that for any phase $A$, $\mu_A^\circ = G^f_A$, the formation Gibbs free energy at standard state.
For a specific phase, one substitutes in Eq.~\ref{eq:littlewoodpotentialexplicit} the Gibbs free energy $G_A$ with
its chemical potential $\mu_A = \mu_A^\circ + k_BT \ln a_A$.

The construction is implemented in an open-source Python module that adapts the Pourbaix-diagram machinery of \texttt{pymatgen}~\cite{Ong2013} to the generalized Littlewood potential.
Each phase is mapped onto the plane coefficients of Eq.~\ref{eq:halfspace}, and the lower envelope and its facets are obtained from a half-space intersection.
For several solute elements, candidate combinations of phases are weighted by balancing their reaction with the reservoir $\{\ce{M}, \ce{X}, \ce{O}\}$ to the target solute composition, and are pre-selected by a convex hull in the space of stoichiometric coefficients and composition, following Ref.~\cite{Patel2019}.
Solid phases are restricted to those stable on the compositional phase diagram at the reservoir chemical potentials.

\subsection*{Finite-temperature free energies}

For a crystalline phase at finite temperature and ambient pressure, we approximate the Gibbs free energy by
\begin{equation}
    G(T) \approx F(T) \approx E_{\mathrm{el}} + F_{\mathrm{vib}}(T)
    ~,
    \label{eq:solidfreeenergy}
\end{equation}
where $E_{\mathrm{el}}$ is the ground-state DFT energy and $F_{\mathrm{vib}}$ is the harmonic vibrational free energy,
\begin{equation}
    \begin{aligned}
        F_{\mathrm{vib}}(T) = & \sum_{\mathbf{q},\nu}\left\{ \frac{\hbar\omega_{\nu}(\mathbf{q})}{2}\right.\\
        + & \left. k_BT\ln\left[1-\exp\left(-\frac{\hbar\omega_{\nu}(\mathbf{q})}{k_BT}\right)\right] \right\}.
    \end{aligned}
    \label{eq:harmonicfreeenergy}
\end{equation}
The frequencies are obtained from DFPT~\cite{Baroni2001} or finite displacements~\cite{Togo2015,Togo2023a, Bastonero2024}.
Equation~\ref{eq:solidfreeenergy} neglects thermal expansion, electronic entropy, configurational entropy, and anharmonicity.
These contributions can be added in the future; for instance, using non-perturbative methods such as the stochastic self-consistent harmonic approximation that can treat strongly anharmonic crystals~\cite{Errea2014,Monacelli2021a}, whose adoption at large scale with first-principles potentials has been made computationally feasible~\cite{Bastonero2026}.

Reference gas phases are treated in the ideal-gas approximation at pressure $p$.
For one molecule,
\begin{equation}
    G(T,p)=H(T)-TS(T,p),
\end{equation}
where the enthalpy contains the electronic energy, zero-point energy, and integrated translational, rotational, and vibrational heat capacities~\cite{Cramer2014}.
The entropy is decomposed as
\begin{equation}
    S=S_{\mathrm{trans}}+S_{\mathrm{rot}}+S_{\mathrm{vib}}+S_{\mathrm{elec}}.
\end{equation}
For a monatomic gas,
\begin{equation}
    S_{\mathrm{trans}}(T,p)=k_B\left\{
      \ln\left[\left(\frac{2\pi m k_BT}{h^2}\right)^{3/2}\frac{k_BT}{p}\right]
      +\frac{5}{2}\right\},
\end{equation}
and $H=E_{\mathrm{el}}+5k_BT/2$ when electronic excitations are neglected~\cite{Reichl2016}.
For a linear molecule in the high-temperature rigid-rotor limit,
\begin{equation}
    \begin{aligned}
    S_{\mathrm{rot}}&=k_B\left[\ln\left(\frac{8\pi^2I k_BT}{\sigma h^2}\right)+1\right],\\
    S_{\mathrm{elec}}&=k_B\ln(2S+1),
    \end{aligned}
\end{equation}
where $I$ is the moment of inertia, $\sigma$ is the rotational symmetry number, and $S$ is the total electronic spin.
For a homonuclear diatomic molecule with atomic mass $m$ and bond length $a$, $I=ma^2/2$ and $\sigma=2$.
Thus \ce{O2} is treated with $S=1$, whereas closed-shell \ce{Cl2} and \ce{F2} have $S=0$.
The vibrational term is evaluated as a harmonic molecular mode.
In practice, all these contributions are evaluated consistently with the \texttt{IdealGasThermo} implementation described below.

Liquid and dissolved specie free energies are not explicitly calculated.
In fact, their calculation is notoriously more difficult and computationally expensive than for previous systems, typically requiring extensive molecular dynamics trajectories and the use of techniques such as thermodynamics integration or particle insertion~\cite{Frenkel2011,Widom1963,Liu2014,Song2017,Song2017a,Bonnet2024}, and deserves a dedicated study.
Here, we approximate the molten-salt free energy by the harmonic free energy of its lowest-energy stoichiometric crystal at the temperature of interest.
This approximation is motivated by the equality of solid and liquid Gibbs free energies at melting and is intended primarily to capture comparative chemical trends near the melting range, but holds surprisingly well even at temperatures above the melting point, as shown in Supplementary Note~1.
For a dissolved species, its standard chemical potential is approximated by the free energy of the stoichiometrically corresponding halide at unit activity (for example, \ce{FeCl2} for \ce{Fe^2+}), and then assuming ideal dilution to the actual activity of the species in solution.
This approximation was also adopted in the original Littlewood construction~\cite{Littlewood1962} and in other works~\cite{Yan2009,Ham2026}.

\subsection*{Reference-state calibration}
Semi-local density functionals exhibit systematic errors in molecular reference energies~\cite{Perdew1981}. 
Following established first-principles thermochemistry strategies~\cite{Hammer1999,Wang2006}, we apply element-specific rigid shifts to the oxygen and halogen gas reference energies.
The shifts are fitted by least squares to the experimental NIST-JANAF~\cite{Chase1998} formation free energies of the alkali and alkaline-earth oxides and halides over 300--1500~K, yielding $\delta_\mathrm{Cl} = +0.4224$, $\delta_\mathrm{F} = +0.3773$, and $\delta_\mathrm{O} = -0.0265$~eV/atom and reducing the deviation from experiment to below 0.1~eV per formula unit for the halides (Supplementary Fig.~2 and Supplementary Note~1).
We avoid including in the fit other compounds (e.g., ternay oxides or oxyhalides) to maximize the accuracy of the formation energy of the salts, so to have accurate electrochemical stability windows.

\subsection*{Computational details}
The structures used to construct the first-principles phase diagrams are taken from a curated experimental database containing more than 80\,000 inorganic crystals, as described in Ref.~\onlinecite{Huber2026}.
Ground-state total energies are computed using DFT, while the temperature-dependent free energies are obtained within the harmonic approximation from dynamical matrices calculated with DFPT or finite differences.
The DFT calculations are performed using the \textsc{Quantum ESPRESSO} distribution v7.2 and are automatically managed through the \texttt{aiida-quantumespresso}, \texttt{aiida-quantumespresso-ph}, and \texttt{aiida-vibroscopy}~\cite{Bastonero2024} plugins of AiiDA~\cite{Huber2020,Uhrin2021}.
Spanning the chemical space of O, Si, Al, Li, Mg, Ca, Na, K, Cl, and F, we performed variable-cell geometry optimizations of 1014 crystal structures containing up to 100 atoms.
For all the 1014 relaxed structures, the harmonic phonons are obtained by finite differences using \textsc{Phonopy}~\cite{Togo2015, Togo2023, Togo2023a}, automated through the \texttt{aiida-phonopy} plugin~\cite{Bastonero2024}, with the fine-tuned machine-learning interatomic potential.
For a subset of 235 of these structures, containing up to 64 atoms, we further computed the dynamical matrices from DFPT: 166 of them span the O--Si--Ca--Cl chemical space of the Littlewood diagrams of Fig.~\ref{fig:yasuda}, while the remaining 69 are the alkali and alkaline-earth halides and oxides on which the reference states are calibrated (see also Supplementary Note~1).
The elemental reference states require the \ce{O2}, \ce{Cl2}, and \ce{F2} molecules, whose total energies and vibrational frequencies we obtain by finite differences in a periodic box, as well as the total energies of the isolated atoms of the metallic and semimetallic elements, which enter the elemental references as their monatomic vapour and thus capture the transition to the gas phase of the most volatile metals at high temperature (e.g., potassium evaporates at 1033~K).
Whenever several phases share the same stoichiometry, namely polymorphs, or a crystal and its monatomic or molecular gas, we take at each temperature the one with the lowest Gibbs free energy.
This applies to the elemental references, for which the harmonic crystal and the monatomic vapour at 1~bar are compared, to the molten-salt and cation-oxide reservoirs, and to the gas species entering the diagrams.
The solute phases instead enter the construction individually, which also gives, for any phase, its distance from the lower envelope at given $(\phi, \ce{pO^{2-}})$, a measure of its electrochemical metastability analogous to the decomposition energy of aqueous Pourbaix diagrams~\cite{Singh2017}.

The relevant computational details are summarized below.

\subsubsection*{Density-functional theory (DFT)}
We use the PBEsol  flavor~\cite{Perdew2008} for the exchange-correlation functional, and employ 
pseudopotentials from the SSSP  PBEsol  precision library (version 1.3)~\cite{Prandini2018, 
Garrity2014, Schlipf2015, Willand2013, Corso2014, Topsakal2014, Setten2018}.
The Brillouin zone is sampled with a $\mathbf{k}$-point distance of $\Delta\mathbf{k}=0.2$~\AA$^{-1}$.
A cold smearing~\cite{Marzari1999} is used with a broadening parameter of 0.01\,Ry.
The kinetic-energy cutoffs for the expansion of KS wavefunctions are set to the 
recommended values of the SSSP library~\cite{Prandini2018,Bosoni2023}. 
The crystal structures are optimized using the Broyden-Fletcher-Goldfarb-Shanno (BFGS) 
algorithm, with a convergence threshold for the total energy of $10^{-6}\,$Ry/atom, 
for forces of $10^{-4}\,$Ry/Bohr, and for pressure of 0.1~kbar.

The elemental reference states involve isolated systems, namely the single atoms and the \ce{O2}, \ce{Cl2}, and \ce{F2} molecules, which we describe within the same plane-wave framework by placing them in large simulation boxes sampled at the $\Gamma$ point only.
The atoms are placed at the center of a fixed cubic box of 30~\AA{} side, whereas for the molecules we retain at least 15~\AA{} of vacuum along each direction and increase the kinetic-energy cutoff to 110~Ry.
For the molecules, the spurious electrostatic interaction with the periodic replicas is removed by correcting both the total energy and the self-consistent potential following the Martyna-Tuckerman scheme~\cite{Martyna1999}, as implemented in \QE.

\subsubsection*{Density-functional perturbation theory (DFPT)}
Density-functional perturbation theory (DFPT) calculations to obtain the dynamical matrix, and the dielectric and Born effective charge tensors (for insulators only), were performed using the \textsc{PHonon} code~\cite{Baroni2001} of \QE.
A $\mathbf{q}$-point distance of $\Delta\mathbf{q}=0.4$~\AA$^{-1}$ is chosen, i.e., twice the $\mathbf{k}$-point distance used for the ground-state calculations.
A tight threshold of $\texttt{tr2\_ph} = 10^{-20}$, controlling the self-consistent convergence of the Sternheimer equations, is used to achieve accurate results.

For the molecules, vibrational frequencies are computed by finite differences at the $\Gamma$ point employing the Martyna-Tuckerman scheme~\cite{Martyna1999}, as explained in the previous section.
A distance of 0.01~\AA{} is used to displace the atoms and compute the force constants matrix.

\subsubsection*{Machine-learned interatomic potential}
\label{sec:fine-tuned-model}
We fine-tuned the NequIP-OAM-L foundation model~\cite{Kavanagh2026} on the MAD dataset~\cite{Mazitov2025a} using the \textsc{NequIP} framework~\cite{Batzner2022,Tan2025}.
From the MAD dataset we removed all structures containing elements not supported by NequIP-OAM-L, as well as all configurations with a maximum force component exceeding 20~eV/\AA; the resulting dataset was randomly split into training (87.5\%), validation (10\%), and test (2.5\%) sets.
We minimized a combined energy--force loss function based on per-atom energies, with initial coefficients of $1/6$ and $5/6$, respectively, adaptively rebalanced during training using the SoftAdapt scheme~\cite{Heydari2019} ($\beta = 0.01$, updated every 5 epochs).
The optimization used the Adam optimizer~\cite{Kingma2015} with the AMSGrad variant~\cite{Reddi2018}, a learning rate of 0.001, a batch size of 4, and an exponential moving average of the model weights; the learning rate was reduced by a factor of 0.8 upon plateaus of the validation loss (patience of 40 epochs, relative threshold of 0.01, minimum learning rate of $10^{-6}$).
Training ran for at most 3000 epochs within a 24-hour wall-time on 10 NVIDIA GH200 nodes, with early stopping on the validation force mean absolute error (patience of 50 epochs); the model with the lowest validation loss was retained.
The per-type energy scales were initialized to the per-type root-mean-square forces of the dataset and kept fixed, while the trainable per-type energy shifts were initialized to zero, since atomic reference energies are already removed in the MAD dataset~\cite{Mazitov2025a}.
A radial cutoff of 6~\AA{} was employed, consistent with the foundation model.

Phonon calculations with NequIP-OAM-MAD-L were performed using the finite-displacement method as implemented in \textsc{Phonopy}~\cite{Togo2015,Togo2023,Togo2023a}, using displacement distance of 0.01~\AA{} , employing supercells commensurate with the corresponding DFPT $\mathbf{q}$-point grids where available, and otherwise determined by the same $\Delta\mathbf{q} = 0.4$~\AA$^{-1}$ criterion.
The submission of all calculations was orchestrated through the \texttt{aiida-phonopy} package~\cite{Bastonero2024}, using the \texttt{PhonopyAseWorkChain} workflow.

\subsubsection*{Free energies}
To obtain the final free energies of solids, the dynamical matrices are interpolated on a fine mesh ($\Delta\mathbf{q} \approx 0.02$~\AA$^{-1}$) using \textsc{Phonopy}~\cite{Togo2015,Togo2023,Togo2023a}, considering the dipole-dipole correction in the long-wavelength limit for insulators~\cite{Giannozzi1991,Gonze1994,Gonze1997,Baroni2001} (only applied to the DFPT calculations, not to the finite-difference ones using the finte-tuned MLIP).
Imaginary harmonic frequencies, reported as negative, are excluded from the sum of Eq.~\ref{eq:harmonicfreeenergy} following the default convention of \textsc{Phonopy}, so that a structure with unstable modes contributes the free energy of its stable modes only.
We flag a structure as dynamically unstable when any frequency on a dense $\mathbf{q}$-mesh lies below $-0.1$~THz ($-3.3$~cm$^{-1}$), a tolerance that absorbs the acoustic-sum-rule noise at $\Gamma$.
The minimum frequency and the number of imaginary modes are flagged on every calculation.

For the free energy of gas molecules and atoms in the ideal gas limit, we use the \texttt{thermochemistry}
module of the Atomic Simulation Environment (ASE)~\cite{HjorthLarsen2017} Python
package, specifically the \texttt{IdealGasThermo} class.
We consider all the possible entropy terms: electronic, vibrational (harmonic approximation),
translational, rotational, and spin.

\section*{Data availability}
The data used to produce the results of this work will be mae available in the Materials Cloud Archive~\cite{Talirz2020}.

\section*{Code availability}
All the code developed and presented here is open source and will be made available on GitHub, and adapts the Pourbaix-diagram module of \texttt{pymatgen}~\cite{Ong2013,Patel2019}.
The AiiDA infrastructure~\cite{Pizzi2016, Huber2020, Uhrin2021} (\href{https://github.com/aiidateam/aiida-core}{https://github.com/\-aiidateam/\-aiida-core}), the \texttt{aiida-quantumespresso} plugin~\cite{Huber2021,Huber2026} (\href{https://github.com/aiidateam/aiida-quantumespresso}{https://github.com/\-aiidateam/\-aiida-quantumespresso}), the \texttt{aiida-quantumespresso-ph} plugin (\href{https://github.com/aiidateam/aiida-quantumespresso-ph}{https://\-github\-.com/\-aiidateam/\-aiida-\-quantumespresso-ph}), the \texttt{aiida-vibroscopy} package (\href{https://github.com/bastonero/aiida-vibroscopy}{https://\-github\-.com/\-bastonero/\-aiida-\-vibroscopy}), and the \textsc{Quan\-tum ES\-PRES\-SO} code suite~\cite{Giannozzi2009, Giannozzi2017, Giannozzi2020} (\href{https://gitlab.com/QEF/q-e}{https://\-gitlab.com/\-QEF/\-q-e}), are all open-source available software.

\section*{Acknowledgements}
L.B., L.C.C., and N.M. gratefully acknowledge support from the Deutsche Forschungsgemeinschaft (DFG) under Germany’s Excellence Strategy (EXC 2077, No. 390741603, University Allowance, University of Bremen) through the ``U Bremen Excellence Chair Program''.
N.M. also acknowledges support by the NCCR MARVEL, a National Centre of Competence in Research, funded by the Swiss National Science Foundation (Grant number 205602).
Computer time was provided by the Swiss National Supercomputing Centre (CSCS) under project No. lp18.

\section*{Competing Interests}
The authors declare no competing interests.

\putbib[biblio]
\end{bibunit}

\clearpage
\onecolumngrid
\setcounter{figure}{0}
\renewcommand{\figurename}{Supplementary Figure}
\renewcommand{\thefigure}{\arabic{figure}}
\setcounter{table}{0}
\renewcommand{\tablename}{Supplementary Table}
\setcounter{section}{0}

\begin{bibunit}[apsrev4-2]

\begin{center}
    {\large\bfseries Supplementary Information for \\[4pt] ``A grand potential for molten-salt electrochemical phase diagrams from first principles''}\\[10pt]
    Lorenzo Bastonero\,\orcidlink{0000-0001-9374-1876},$^{1,\,*}$ Lucio Colombi Ciacchi\,\orcidlink{0000-0003-1444-9733},$^{1,\,2}$ and Nicola Marzari\,\orcidlink{0000-0002-9764-0199}$^{1,\,3,\,4}$\\[8pt]
    {\small\itshape
    $^{1}$\BREMEN\\
    $^{2}$\HMI\\
    $^{3}$\THEOSMARVEL\\
    $^{4}$\PSI\\[4pt]
    }
    {\small $^{*}$lbastone@uni-bremen.de}
\end{center}

\section{Supplementary Notes}
\renewcommand{\thesubsection}{\arabic{subsection}}
\makeatletter\renewcommand{\p@subsection}{}\makeatother

\subsection{Calibration of the gas reference energies}
\label{note:calibration}
An element-specific rigid shift $\delta_X$ of the gas reference corresponds to correcting each formation free energy as $G^f \rightarrow G^f - n_X\,\delta_X$, with $n_X$ the number of O or halogen atoms per formula unit; it is equivalent to shifting the total energy of the \ce{X2} molecule by $2\delta_X$ and affects only compounds containing element $X$.
The shifts are determined by a global least-squares fit, $\delta_X = \sum_i n_{X,i}\, d_i / \sum_i n_{X,i}^2$, where $d_i = G^f_{i,\mathrm{theo}} - G^f_{i,\mathrm{exp}}$ runs over all temperature--compound points.
More specifically, the NIST-JANAF crystal-phase formation free energies of the following compounds, sampled every 50~K between 300 and 1500~K (125 points per element), are used in the fit:
\begin{itemize}
    \item Cl: \ce{LiCl}, \ce{NaCl}, \ce{KCl}, \ce{MgCl2}, \ce{CaCl2}, \\
    \item F: \ce{LiF}, \ce{NaF}, \ce{KF}, \ce{MgF2}, \ce{CaF2}, \\
    \item O: \ce{Li2O}, \ce{Na2O}, \ce{K2O}, \ce{MgO}, \ce{CaO}.
\end{itemize} 
The theoretical formation energies are fully ab initio: PBEsol electronic energies of the relaxed solids combined with harmonic vibrational free energies (zero-point energy included) from density-functional perturbation theory, taking the lowest-energy polymorph, referenced to the elemental solids and to the \ce{O2}, \ce{Cl2}, and \ce{F2} molecules computed at the same level; the resulting shifts are then applied consistently to both the DFPT- and the ML-based diagrams.
The fit yields $\delta_\mathrm{Cl} = +0.4224$, $\delta_\mathrm{F} = +0.3773$, and $\delta_\mathrm{O} = -0.0265$~eV/atom, reducing the root-mean-square deviation per formula unit from 0.567 to 0.091~eV for the fluorides and to 0.060~eV for the chlorides. 
For the oxides the shift is marginal and the deviation remains at 0.138~eV.

In Supplementary Fig.~\ref{fig:salt_formation_energies} we compare theoretical free energies describe harmonic crystal with the experimental data between 700 and 1500~K, with and without the shifts.
The comparison extends through the experimental melting points, indicated by a vertical line when occurring (we note that not all compounds have melting points within this range).

We note that the shifts are fitted on crystalline salts and oxides, and applied to condensed phases only.
A wide range of options of such corrections have been proposed in the past~\cite{Wang2006,Grindy2013}, more recently even applied depending on the bonding environment of the anion~\cite{Friedrich2019}.
A similar practice of the one employed here, i.e., correcting only condensed phases free energies alongside ``uncorrected'' or experimental molecular species, underlies the computational Pourbaix diagrams of Ref.~\onlinecite{Persson2012}.
As such, molecular species are included in the phase diagram construction as ideal gases at their standard state (see Methods), without the reference shift.

\subsection{Validation of the fine-tuned machine-learning potential}
\label{note:mlip_validation}
The fine-tuning set comprises the MAD structures containing neither Po nor Rn and whose
largest force component does not exceed 20~eV/\AA{} (83,835 structures of the MAD dataset~\cite{Mazitov2025a},
spanning 83 elements from H to Bi), randomly partitioned into training, validation and
test subsets with fractions 0.875/0.1/0.025, i.e. 73,356/8,384/2,095 structures.
The reference energies and forces of MAD were computed with PBEsol and the SSSP efficiency pseudopotentials (version 1.2) with a denser $k$-point sampling, whereas our DFT and DFPT calculations use SSSP precision (version 1.3); both libraries are selected by verification against all-electron benchmarks~\cite{Prandini2018,Bosoni2023}, so that for the elements of this work the two are expected to be closely consistent.
Moreover, the fine-tuned potential supplies only the vibrational term, so the static energies of the diagrams are computed throughout at our settings, and any residual mismatch in the force constants is part of what the DFPT benchmark of Supplementary Note~\ref{note:ml_phonons} measures.
To quantify its accuracy we evaluate the model on all three of its own splits and
compare the predictions to the reference values of the dataset.
The resulting errors are summarized in Supplementary Table~\ref{tab:mlip_metrics}, and
the parity of the per-atom energies and of the force components on the validation and test
splits is shown in Supplementary Figure~\ref{fig:mlip_parity}.

The mean absolute errors of the three splits agree to within 3.2~meV/atom for the energies and 3.3~meV/\AA{} for the forces, with no degradation from the training to the validation/test splits.
Since the splits are random partitions of one dataset, this rules out memorization of the training set, but does not by itself establish transferability to structures outside it, which we assess instead against DFPT in Supplementary Note~\ref{note:ml_phonons}.
The energy RMSE of the validation split is an outlier-dominated quantity: two
near-equilibrium 36-atom transition-metal structures (Fe$_{10}$W$_{24}$C$_2$ and
Ti$_{24}$Fe$_{10}$O$_2$, with per-atom energy errors of 10.1 and 8.5~eV) carry almost
all of it, and removing the single worst structure already reduces the validation energy
RMSE from 0.1517 to 0.1038~eV/atom, and removing the five worst to 0.0423~eV/atom.
The force errors are insensitive to such removals, so the force metrics are robust as
reported.
Restricting the analysis to the chemistries relevant to the present work, i.e. to the structures built exclusively from O, Si, Al, Li, Mg, Ca, Na, K, Cl and F, the two held-out splits contain 267 structures, for which we obtain an energy MAE (RMSE) of 0.0349 (0.0559)~eV/atom and a force MAE (RMSE) of 0.0562 (0.1007)~eV/\AA{}: the forces, which are the quantity of interest here, are predicted more accurately than for the dataset as a whole.

\subsection{Accuracy of the machine-learned vibrational free energies}
\label{note:ml_phonons}
We compare the vibrational contribution to the Gibbs energy of the solid phases between the fine-tuned MLIP finite difference phonons and density-functional perturbation theory (DFPT) calculations.
To quantify the error, we compare the two approaches on the subset for which both are available, namely the 235 structures of the DFPT calculations.
Both calculations start from the same DFT-relaxed geometry and share the same thermodynamic post-processing.
They differ in one respect: the non-analytical corrections to the dynamical matrices of insulators require the dielectric tensor and the Born effective charges~\cite{Baroni2001,Togo2015,Bastonero2024}, which DFPT provides and the machine-learned potential does not, so the machine-learned phonons are interpolated without these corrections.
The comparison therefore measures the combined effect of the machine-learned force constants and of the omitted non-analytical corrections, which is the error we would commit when applying the pipeline at scale.
As shown in Fig.~\ref{fig:vib_ml_vs_dfpt}, the agreement is excellent at the temperatures relevant to molten-salt operation: the mean absolute error is 5.6~meV/atom at 500~K, 12.7~meV/atom at 1000~K and 20.7~meV/atom at 1500~K, with root-mean-square errors of 10.3, 25.8 and 44.1~meV/atom, respectively.
The error grows linearly with temperature, consistent with a residual error in the phonon spectrum entering through $-TS_\mathrm{vib}$.

\subsection{Dynamical stability of the structures}
\label{note:stability}
All diagrams of the main text include every structure of the dataset, including those with imaginary harmonic frequencies, which are omitted from the vibrational sum (see Methods).
Here we show the constructions restricted to dynamically stable structures, flagged on a dense $\mathbf{q}$-mesh with a $-0.1$~THz tolerance.
We identified two causes underlying the disagreements between DFPT and the MLIP on the same structure.
First, the force constants of insulators computed using the foundation MLIP are not interpolated using the non-analytical corrections.
Focusing on the set of structures employed for the \ce{Ca}--\ce{Si}--\ce{O}--\ce{Cl} system in Fig.~2 of the main manuscript, employing the DFPT Born charges and dielectric tensor restores the stability of 7 of the 24 structures, and shrinks the imaginary modes of most of the others.
Second, a residual long-wavelength error of the potential on silica and silicates remains for 17 structures, and for the metal \ce{Ca3SiO}, whose low-frequency zone-center modes of DFPT become imaginary with the MLIP.
Imaginary modes of the latter type (i.e., at the $\Gamma$ point) may also be related to the use of the same structure optimized with DFT, meaning that the structure is not at its minimum when computing phonons with the MLIP.
Lastly, there are also structures showing a ``genuine'' harmonic instability, as for \ce{KAlO2}, flagged by both methods in all polymorphs, and plausibly for \ce{Ca3Si(ClO2)2} and \ce{K3AlF6}. These soft modes identify phases where the harmonic approximation may not hold anymore, and may be stabilized at high-temperature by thermal fluctuations,requiring an anharmonic treatment of the lattice dynamics.

\clearpage
\section{Supplementary Figures}
\begin{figure*}[h!]
    \centering
    \includegraphics[width=\textwidth]{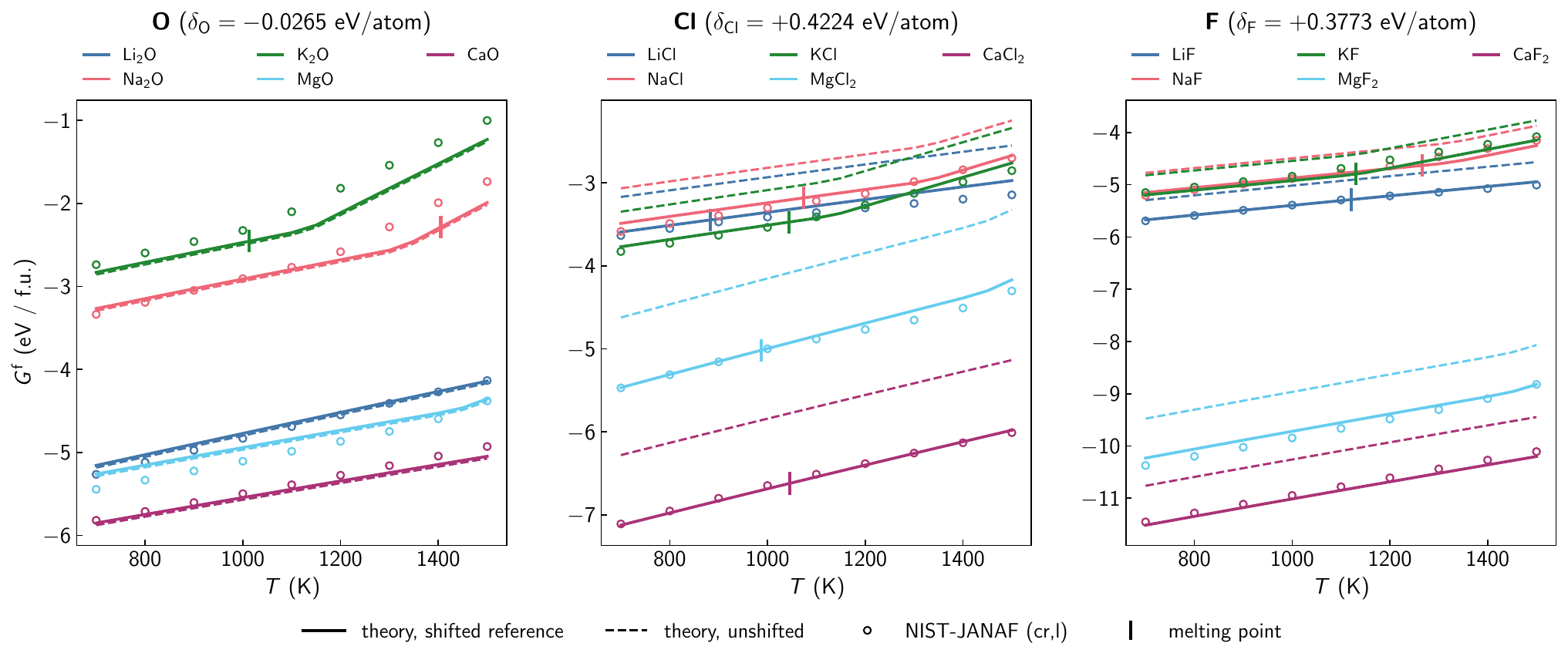}
    \caption{
        \textbf{Calibration of the gas references against experimental formation energies.} 
        Gibbs formation energies $G^f$ (per formula unit) of the alkali and alkaline-earth oxides (left), chlorides (center), and fluorides (right) between 700 and 1500~K.
        Solid lines: theoretical values with the calibrated rigid shift $\delta$ applied to the corresponding gas reference (\ce{O2}, \ce{Cl2}, \ce{F2}; per-atom values in the panel titles); dashed lines: without shift. 
        Open circles: NIST-JANAF experimental values (crystal and liquid tables); vertical ticks mark the melting points. 
        Note that solid and dashed lines nearly coincide in the O panel ($\delta_\mathrm{O} = -0.0265$~eV/atom). 
        See Supplementary Note~\ref{note:calibration} for the fitting procedure.
    }
    \label{fig:salt_formation_energies}
\end{figure*}
\begin{figure*}[h!]
  \centering
  \includegraphics[width=0.85\textwidth]{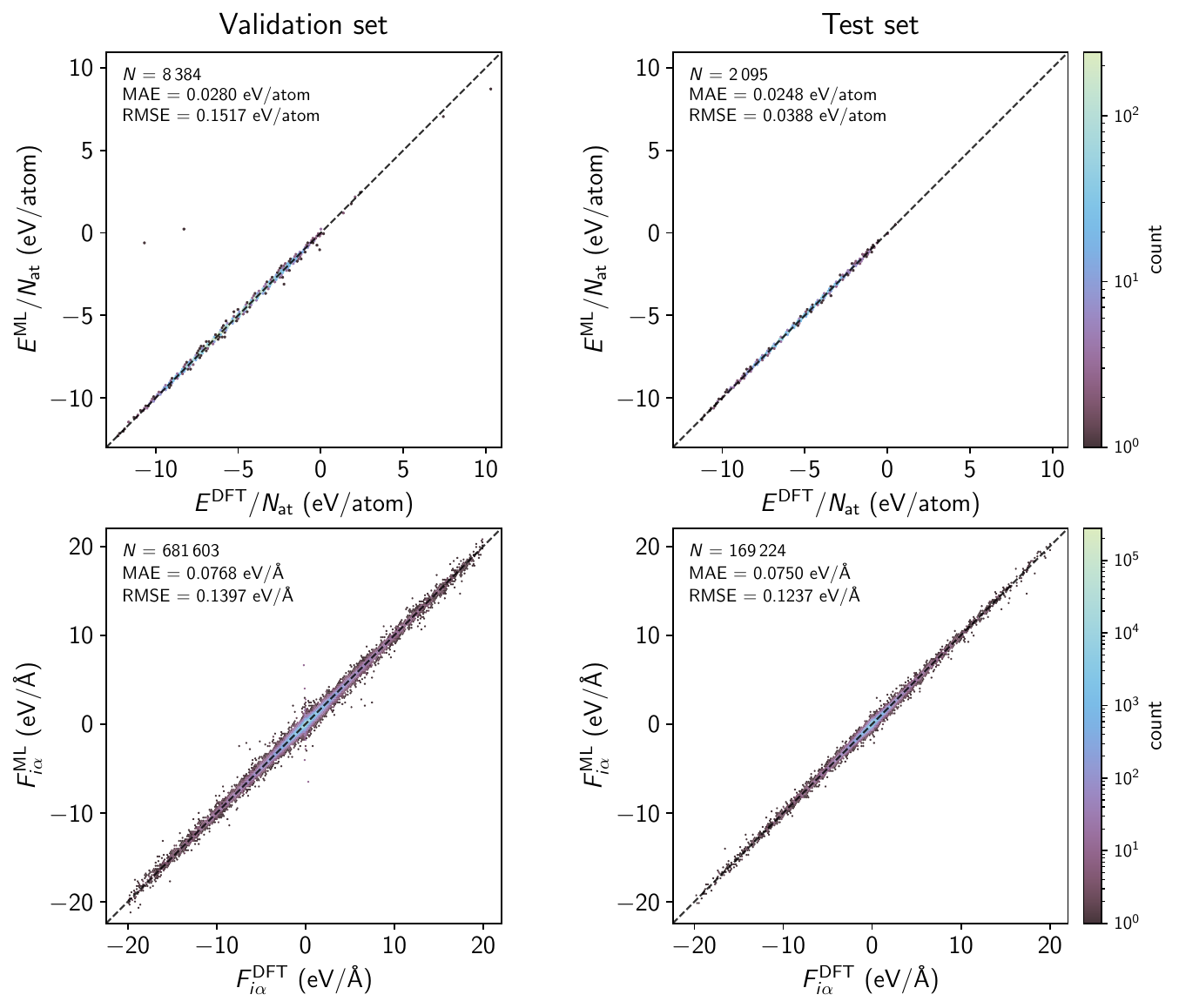}
  \caption{%
    \textbf{Parity of the fine-tuned machine-learning interatomic potential on the validation and test splits.}
    Top row: energy per atom; bottom row: Cartesian force components; left column: validation split; right column: test split.
    The color encodes the number of data points per hexagonal bin on a logarithmic scale, and the dashed line marks perfect agreement; axis limits are shared within each row so that the two splits are directly comparable.
    Each panel reports the number of data points $N$ (structures for the energies, force components for the forces) together with the mean absolute error (MAE) and the root-mean-square error (RMSE) with respect to the corresponding dataset.
    The few points lying visibly off the diagonal in the validation energy panel are two 36-atom transition-metal structures (Fe$_{10}$W$_{24}$C$_2$ and Ti$_{24}$Fe$_{10}$O$_2$) that alone account for almost the entire validation energy RMSE; see Supplementary Note~\ref{note:mlip_validation}.
    \label{fig:mlip_parity}
  }
\end{figure*}
\begin{figure*}[h!]
  \centering
  \includegraphics[width=\textwidth]{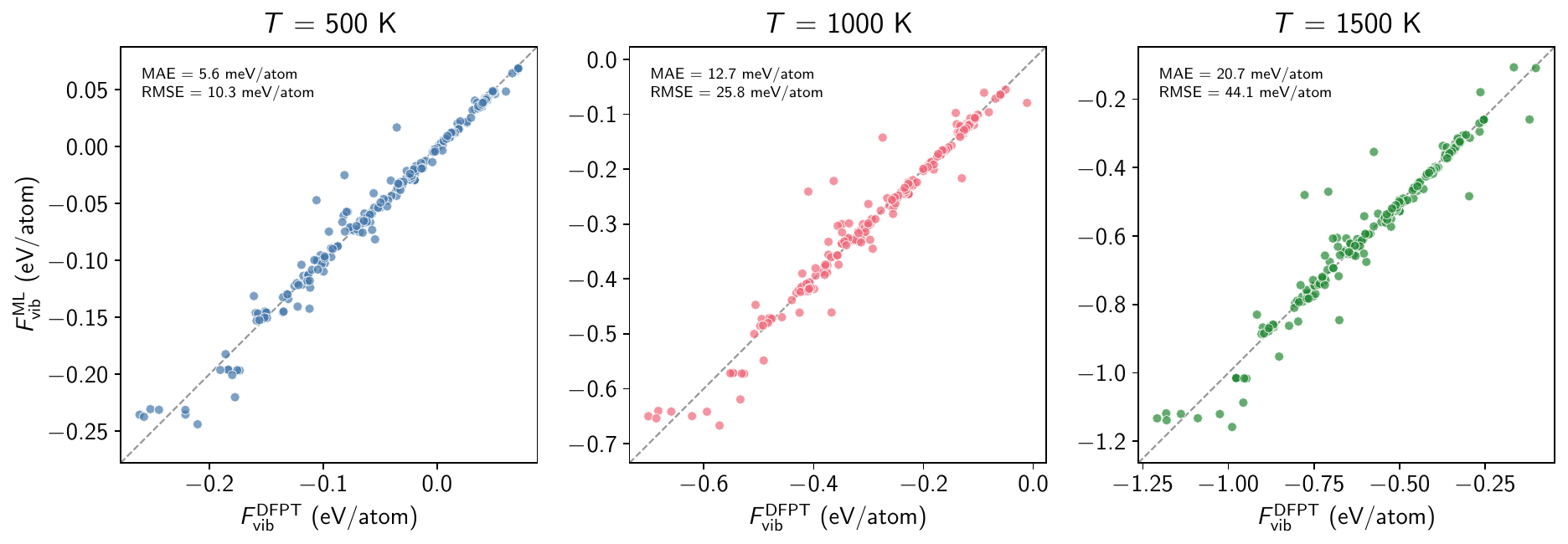}
  \caption{
    \textbf{Harmonic vibrational free energies of the fine-tuned machine-learning interatomic potential compared to density-functional perturbation theory.}
    Parity plots of the vibrational (Helmholtz) free energy per atom, $F_\mathrm{vib}$ (zero-point energy included), at $T=500$, 1000 and 1500~K (left to right), for the 235 crystal structures whose phonons were computed both with DFPT and with the MLIP.
    The dashed line marks perfect agreement; the mean absolute error (MAE) and the root-mean-square error (RMSE) with respect to DFPT are reported in each panel.
}
\label{fig:vib_ml_vs_dfpt}
\end{figure*}
\begin{figure*}[h!]
    \centering
    \includegraphics[width=0.9\textwidth]{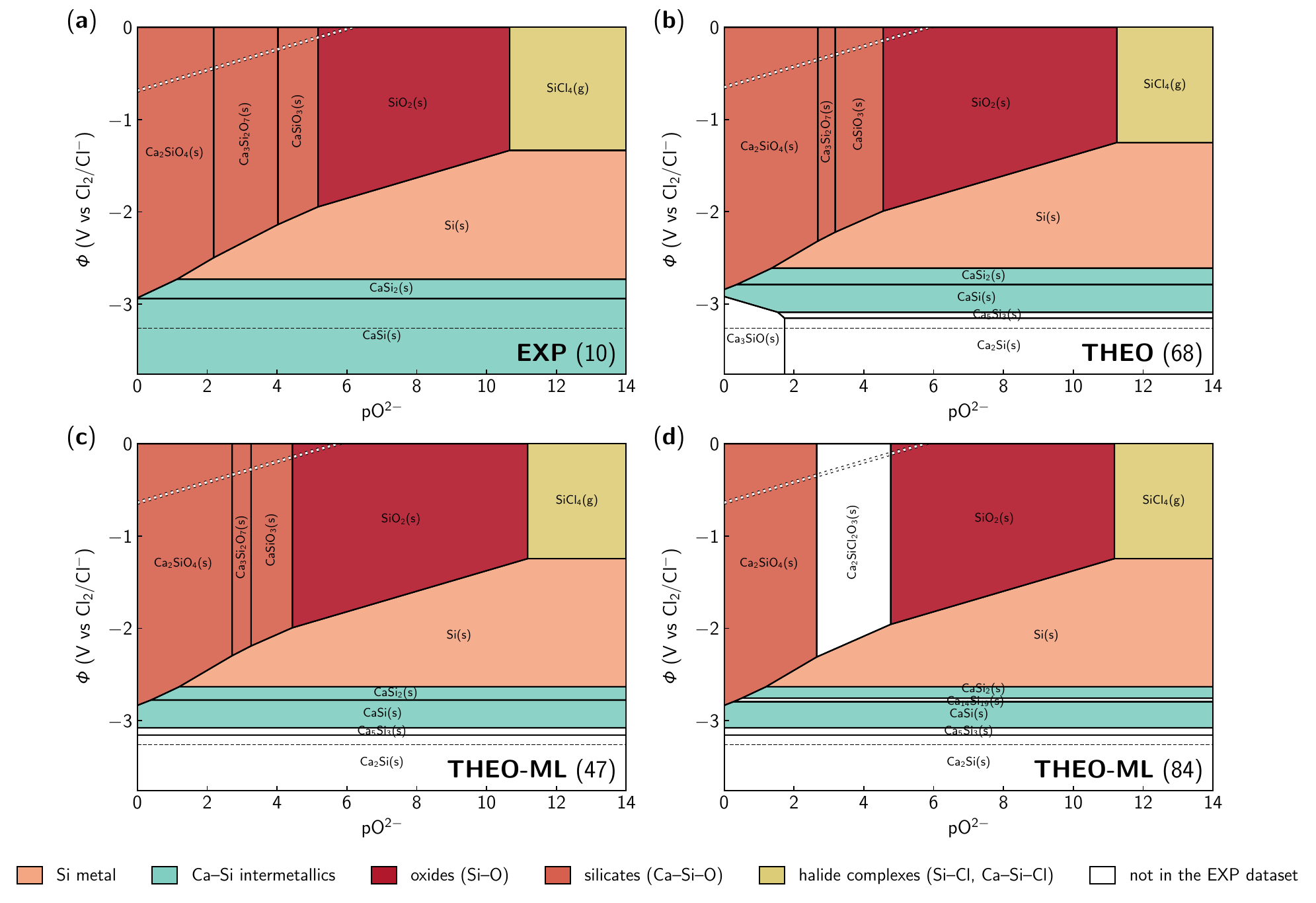}
    \caption{
        \textbf{Littlewood diagrams of silicon in molten \ce{CaCl2} restricted to dynamically stable structures.}
        Same as Fig.~2 of the main text, with the theoretical panels (b)--(d) built only from structures without imaginary harmonic frequencies (see Supplementary Note~\ref{note:stability}); the number in parentheses counts the dynamically stable Si-bearing crystal structures considered, to which gaseous \ce{SiCl4} is added.
    }
    \label{fig:yasuda_stable}
\end{figure*}
\begin{figure*}[h!]
    \centering
    \includegraphics[width=0.96\textwidth]{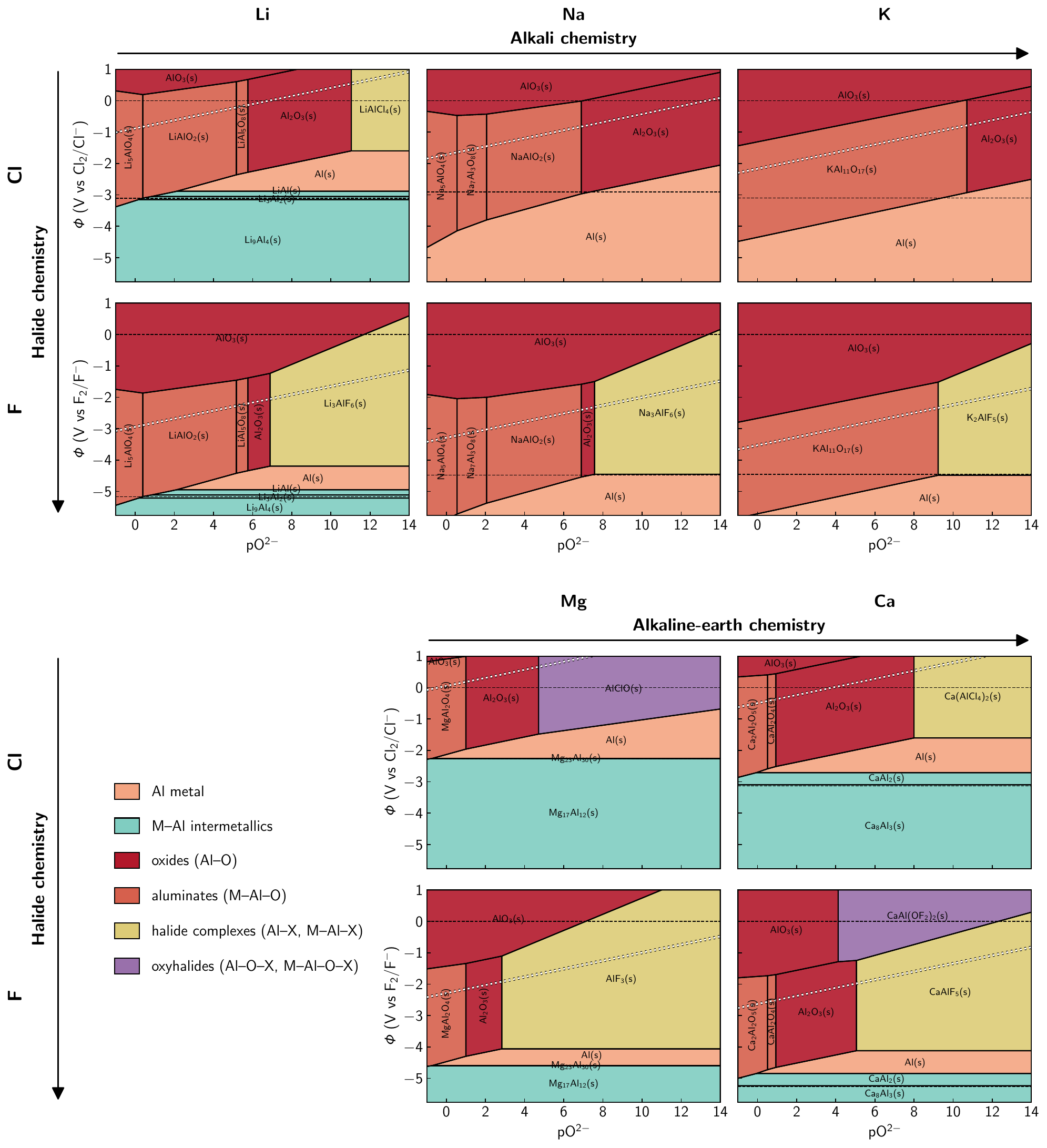}
    \caption{
        \textbf{Generalized Littlewood diagrams of aluminum in molten alkali and alkaline-earth halides restricted to dynamically stable structures.}
        Same as Fig.~3 of the main text, built only from structures without imaginary harmonic frequencies according to the MLIP phonons (see Supplementary Note~\ref{note:stability}).
    }
    \label{fig:al_stable}
\end{figure*}

\clearpage
\section{Supplementary Tables}
\begin{table}[h!]
    \centering
    \caption{
        Accuracy of the fine-tuned MLIP on the training, validation and test splits of its fine-tuning dataset. 
        Energies in meV/atom, forces in meV/\AA{}.
        \label{tab:mlip_metrics}
    }
    \setlength{\tabcolsep}{10pt}
    \renewcommand{\arraystretch}{1.3}
    \begin{tabular}{lrcccc}
        \toprule
        split & structures & \multicolumn{2}{c}{$E/N_{\mathrm{atoms}}$} & \multicolumn{2}{c}{$F$} \\
        \cmidrule(lr){3-4}\cmidrule(lr){5-6}
        {} & {} & MAE & RMSE & MAE & RMSE \\
        \midrule
        training   & 73\,356 & 24.8 & 48.8 & 73.5 & 120.9 \\
        validation &  8\,384 & 28.0 & 151.7 & 76.8 & 139.7 \\
        test       &  2\,095 & 24.8 & 38.8 & 75.0 & 123.7 \\
        \bottomrule
    \end{tabular}
\end{table}

\clearpage
\section{Supplementary References}
\putbib[biblio]
\end{bibunit}

\end{document}